\input{aipcheck}

\documentclass[
    ,final            % use final for the camera ready runs
  ]
  {aipproc}

\layoutstyle{6x9}

\begin{document}

\title{New Results on the $\pi^{+}\pi^{-}$ Electroproduction Cross Sections off Protons}


\author{G.V.~Fedotov}{
  address={University of South Carolina, Columbia, SC 29210, USA}
  ,altaddress={Moscow State University, Skobeltsyn Institute of
Nuclear Physics, 119899 Moscow, Russia}
}

\author{R.W.~Gothe}{
  address={University of South Carolina, Columbia, SC 29210, USA}
}

\begin{abstract}
In these proceedings we present preliminary $\pi^{+}\pi^{-}$ electroproduction cross sections
off protons in the kinematical area of  1.4~GeV~$< W <$~1.8~GeV  and 0.4~GeV$^{2}$~$< Q^{2} < 1.1$~GeV$^{2}$.
Our results extend the kinematical coverage for this exclusive channel with respect to previous measurements.
Furthermore, the $\pi^{+}\pi^{-}$ electroproduction cross sections were obtained for $Q^2$-bins of much smaller size. 
The future analysis of this data within the framework of the JLAB-MSU~reaction model (JM) will considerably improve  
our knowledge on the $Q^2$ evolution of the transition $\gamma_{v}NN^*$ electrocouplings, in particular for the resonances with masses above
1.6 GeV.
\end{abstract}

\maketitle

\section{Introduction}

Measurements of the $\pi^{+}\pi^{-}$ electroproduction cross sections
represent an important part of the efforts in the N$^*$ program with CLAS~\cite{Bu11,Ra11}.
This program is focused on the evaluation of most excited proton  electrocouplings at photon virtualities (Q$^2$) up to 5.0 GeV$^2$,which will
allow us to pin down the active
degrees of freedom in the N$^{*}$ structure at different distance scales, and
to access the non-perturbative strong interaction mechanisms that are responsible
for the formation of the ground and excited nucleon states \cite{CL12,CR11}.
 
 Single and charged double pion exclusive channels are two major contributors
to the meson electroproduction in the N$^{*}$ excitation region with 
different non-resonant mechanisms. A successful description of all
observables in these exclusive channels with consistent  N$^{*}$ electrocouplings
offers evidence for the reliable evaluation of these fundamental
quantities. Moreover, the $\pi^{+} \pi^{-}$ exclusive channel
is very promising for the studies of N$^{*}$s with masses above 1.6 GeV \cite{Bu11,Mo11}. 

Our measurements of $\pi^+\pi^-$ electroproduction cross sections, described in these proceedings,  continue  
the previous studies of this exclusive channel with the CLAS detector \cite{Fedotov08,Ripani}. 
Our preliminary data provid a complementary
kinematical coverage: 1.4~GeV~$< W <$~1.8~GeV  and 0.4~GeV$^{2}$~$< Q^{2} < 1.1$~GeV$^{2}$ in comparison with the previously available measurements, 
and offer more than a factor six smaller
binning over $Q^2$. This kinematical region is suitable to access the electrocouplings and the
$\pi \Delta$, $\rho p$ hadronic decay widths for high lying nucleon resonances as  $S_{31}(1620)$, $S_{11}(1650)$, $F_{15}(1685)$, 
$D_{33}(1700)$, and $P_{13}(1720)$ that have substantial decay probabilities to the $N\pi\pi$ final states \cite{Mo11}.

\section{Data analysis}
Our analysis is focused on the evaluation of the fully integrated and nine one fold differential $\pi^+\pi^-$
electroproduction cross sections off the proton as defined in \cite{Fedotov08}. This
information will allow us to determine $N^*$ electrocouplings within the framework of the reaction 
model JM \cite{Victor,Mo11}.
The analysis is based on the experimental data taken with the CLAS detector during the 2003 e1e run period with the 
electron beam energy 2.039 GeV. The procedure for the extraction of the above mentioned cross sections incorporates: a) the selection
of $e p \pi^{+} \pi^{-}$ exclusive events, and b) the evaluation of detector efficiencies in the five dimensional reaction phase space through
Monte-Carlo simulations. One fold differential cross sections were obtained by integrating the  five fold differential
$\pi^+\pi^-$ electroproduction cross section over all variables except of one of interest, as described in \cite{Fedotov08}.

\begin{figure}[h!]
\label{q2w}
  \includegraphics[height=.25\textheight]{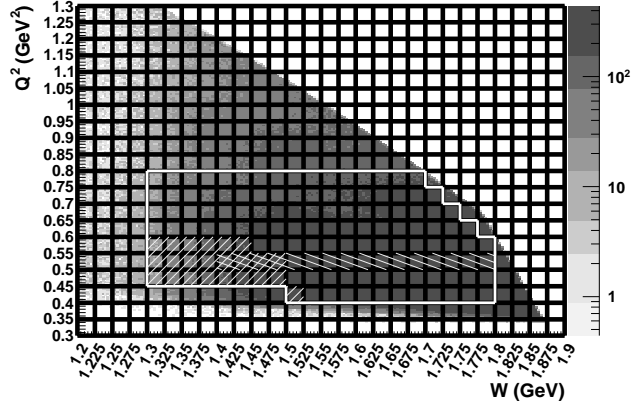}
  \caption{$Q^{2}$ versus $W$ distribution for the selected $e p \rightarrow e'p'\pi^{+}\pi^{-}$ exclusive events. 
  Cross sections were determined inside the area shown by white borders. 
  Hatched areas correspond to the kinematical regions, where the data on $\pi^+\pi^-p$ electroproduction cross sections are available
  from previous CLAS measurements \cite{Fedotov08,Ripani}.}
\end{figure}

\begin{figure}[h!]
  \includegraphics[height=.3\textheight]{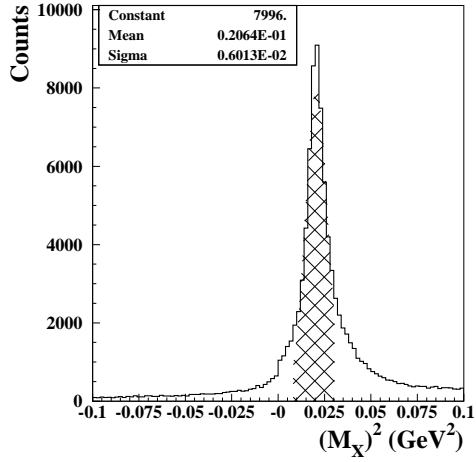}
  \caption{Distribution of the events over the $\pi^-$ missing mass 
  squared that were selected after particle identification and kinematical cuts.}
  \label{mpim}
\end{figure}

The scattered electron was identified by the coincidence of the  electromagnetic calorimeter, Cherenkov counter, 
drift chambers, and time-of-flight detector signals. Positive hadrons (protons and $\pi^{+}$) were identified as particles leaving  hits in  
the drift chambers and time-of-flight scintillators. Further particle identification procedures were similar to those described in 
 \cite{Fedotov08}. The reaction events were selected in four different topologies. In the first one  all final hadrons were detected, while  in the remaning  one of the final hadrons was reconstructed employing energy and momentum
conservation. Maximal statistics is achieved in the missing $\pi^{-}$ topology.   The distribution of the selected events over 
$W$ and $Q^2$ is shown in Fig.~\ref{q2w}. Differential and fully integrated cross sections  were obtained in the 
kinematical region depicted in Fig.~\ref{q2w}  by the white polygon. A considerable extension of the covered kinematical region is achieved in
comparison with  the previous experiments \cite{Fedotov08,Ripani} (hatched areas in Fig.~\ref{q2w}).  
The $\pi^{-}$ missing mass squared distribution of the selected events  is shown in  Fig.~\ref{mpim} and peaks  at the $\pi^{-}$ mass squared.

\section{Preliminary cross sections}

\begin{figure}
\label{wq2}
  \includegraphics[height=.35\textheight]{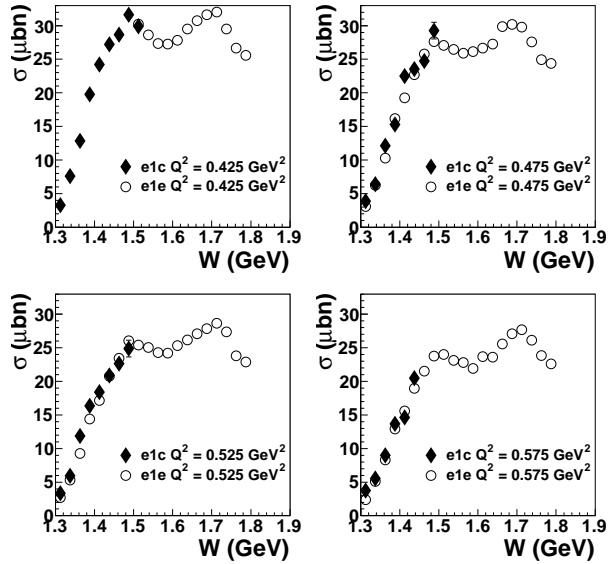}
  \caption{Fully integrated $\pi^+\pi^-$ electroproduction cross sections obtained in our analysis (open circles) in comparison
  with the previously available data \cite{Fedotov08} (filled diamonds).}
\end{figure}

Preliminary results on the fully integrated $\pi^+\pi^-$ electroproduction cross sections obtained in our analysis are shown in 
Fig.~\ref{wq2}  in comparison, with the  previously available data \cite{Fedotov08}.  
The cross sections obtained in the current analysis are in a good agreement  with previously
available results in the overlaping area of $Q^{2}$ and $W$. 
%The uncertainties related to the cross section normalization were estimated from comparison between the elastic
%electron scattering criss sections, obtained for our data set, and the parametrization \cite{Bosted95} of the world data.
%In this way we obtained aprr 5\% systematical unceratainties for fully integrated cross sections. 
The $Q^2$-evolution of the
$\pi^+\pi^-$ electroproduction cross section determined in the entire range of photon virtualities covered by our measurement is depicted in Fig. 4. 

The $W$-dependencies of the fully integrated cross sections show resonant structures in the second and third resonance
regions for all $Q^2$-bins. Therefore, our preliminary data demonstrate a good prospect for the extraction of resonance 
electrocouplings and $\pi \Delta$, $\rho p$ hadronic decay widths employing the JM reaction model \cite{Mo11,Victor}.

\begin{figure}
\label{q2deo}
  \includegraphics[height=.25\textheight]{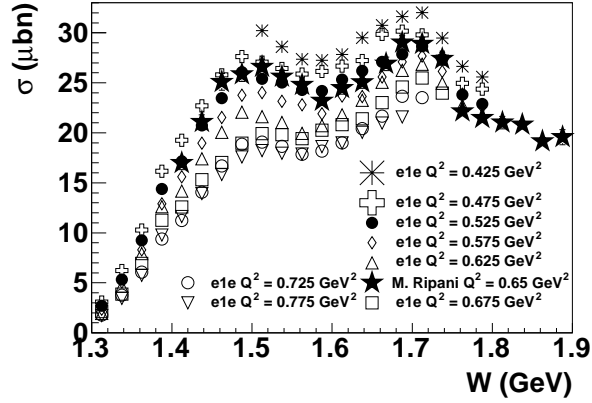}
  \caption{$Q^2$-evolution of the fully integrated $\pi^+\pi^-$ electroproduction cross sections 
  obtainbed in our analysis (all open symbols and black bullets). Previously available data \cite{Ripani} at $Q^{2} = $ 0.65 GeV$^{2}$ are shown as black stars.}
\end{figure}

\section{Conclusions}
In our analysis of the CLAS e1e data obtained  $\pi^+\pi^-p$ electroproduction cross sections extend considerably the kinematical 
range covered in
previous measurements of this exclusive channel. Preliminary results on  nine differential and integrated $\pi^{+}\pi^{-}$p electroproduction 
cross sections were obtained in the kinematic region of 1.4~GeV~$< W <$~1.8~GeV  and 0.4~GeV$^{2}$~$< Q^{2} < 1.1$~GeV$^{2}$, and they are in 
a good agreement with previously available results.
The cross sections show clear indication of 
resonance contributions to the second and third resonance regions, offering  encouraging prospects for the extension of 
our knowledge on $N^*$
electrocouplings, in particular for the resonances with masses above 1.6 GeV.

\bibliographystyle{aipproc}   % if natbib is available

\bibliography{sample}

\IfFileExists{\jobname.bbl}{}
 {\typeout{}
  \typeout{******************************************}
  \typeout{** Please run "bibtex \jobname" to optain}
  \typeout{** the bibliography and then re-run LaTeX}
  \typeout{** twice to fix the references!}
  \typeout{******************************************}
  \typeout{}
 }

\end{document}